\documentclass[twocolumn,prb,superscriptaddress,floatfix,aps,10pt]{revtex4-2}
\pdfoutput=1

\usepackage[utf8]{inputenc}
\usepackage[caption=false,subrefformat=parens,labelformat=parens]{subfig}
\usepackage{graphicx}
\usepackage{float}
\usepackage{amsmath}
\usepackage{amssymb}
\usepackage{dcolumn}
\usepackage{bm}
\usepackage{xcolor}
\usepackage{xfrac}
\usepackage[colorlinks=true, allcolors=blue]{hyperref}
\usepackage{chemformula}
\usepackage[normalem]{ulem}
\usepackage{siunitx}
\usepackage{multirow}

\begin{document}

\title{Raman spectroscopy of graphite with water as the pressure medium}

\author{K. Perry}
\thanks{These three authors contributed equally to this work.}
\affiliation{Department of Physics, Eastern Illinois University, Charleston, Illinois 61920, USA}
\author{A. T. Roy}
\thanks{These three authors contributed equally to this work.}
\affiliation{Department of Physics, Eastern Illinois University, Charleston, Illinois 61920, USA}
\author{A. R. Parmenter}
\thanks{These three authors contributed equally to this work.}
\affiliation{Department of Physics, Eastern Illinois University, Charleston, Illinois 61920, USA}
\author{Y. J. Ryu}
\affiliation{Center for Advanced Radiation Sources, University of Chicago, Chicago, Illinois 60637, USA}
\author{V. B. Prakapenka}
\affiliation{Center for Advanced Radiation Sources, University of Chicago, Chicago, Illinois 60637, USA}
\author{J. Lim}
\email{Corresponding author: jlim5@eiu.edu}
\affiliation{Department of Physics, Eastern Illinois University, Charleston, Illinois 61920, USA}

\date{\today}

\begin{abstract}

We report a high-pressure Raman spectroscopy study of a graphite–water mixture using water as the pressure-transmitting medium up to \SI{9.9}{GPa}. In the graphite-rich region, three characteristic Raman features—the $E_{2\mathrm{g}}^{(1)}$ shear mode, the G band ($E_{2\mathrm{g}}^{(2)}$), and the 2D band—were observed and tracked as a function of pressure. The G band exhibits a pronounced blue shift with increasing pressure, indicating enhanced interlayer coupling between graphite planes. In the water-rich region, the librational band and three distinct O–H stretching modes were identified. Notably, above \SI{8}{GPa}, the slope of the pressure dependence decreases relative to the earlier report, likely due to the influence of the water pressure medium, emphasizing the need for further investigation at higher pressures.

\end{abstract}

\maketitle

\section{Introduction} 
High pressure provides a powerful means to synthesize novel materials that are difficult to obtain under ambient conditions, as it directly modifies the system's energy by compressing its volume~\cite{Jackson_lanthanides_2005}. Moreover, certain materials produced under high pressure can remain stable at ambient conditions if they occupy a kinetically metastable region of the energy landscape~\cite{Allan_EnergyLandscapes_2021}. A well-known example is the transformation of graphite into diamond under combined high-pressure and high-temperature conditions~\cite{BOVENKERK_BundyDiamond_1959}. The cold compression of graphite can induce phase transitions to new carbon allotropes, which has been a subject of intense debate for decades due to the sluggish kinetics of the transformation. Unlike the direct transition from graphite to diamond at high temperature, room-temperature compression produces one or more intermediate, or ``post-graphite,'' phases, whose precise structural characteristics remain unclear. The nature of the resulting post-graphite phase remains poorly understood, and even its crystal structure has not been fully determined~\cite{Wang_graphiteRoomXRD_2012}.

A key effect of high pressure is the conversion of graphite's planar \(sp^{2}\) bonding to the three-dimensional \(sp^{3}\) bonding characteristic of diamond-like structures. The use of water as a pressure-transmitting medium may lower the pressure required for this transition. For instance, studies on bilayer graphene have shown that the G and 2D Raman peaks disappear at approximately 37~GPa in the presence of water, indicating the onset of \(sp^{2}\) to \(sp^{3}\) conversion~\cite{Tao_BilayerGraphene_2020}. It is hypothesized that water molecules facilitate the formation of new \(sp^{3}\) bonds by providing oxygen- or hydrogen-containing functional groups that interact with the carbon lattice under high pressure. In contrast, experiments using alternative pressure media, such as silicone oil, do not show this effect at similar pressures.

High-pressure Raman spectroscopy of graphite–water mixtures offers an opportunity to probe these structural changes in detail. In this study, we investigate how the presence of water, acting as a pressure-transmitting medium, influences the high-pressure behavior of graphite and provides insight into its phase transitions.

\section{Methods}
\label{sec:Methods}
To generate high pressure in the graphite–water mixture, diamond anvil cells (DACs) with two opposing diamond anvils (1/3-carat, type Ia) and \SI{0.3}{\milli\metre} culet diameter were employed. A pre-indented stainless steel gasket (from \SI{200}{\micro\metre} to \SI{50}{\micro\metre} thickness) with a \SI{150}{\micro\metre}-diameter hole drilled at the center using a laser drilling machine was placed between the diamond anvils to form the sample chamber. The graphite samples consisted of thin foils of Highly Oriented Pyrolytic Graphite (HOPG, Alfa Aesar) prepared via the scotch tape method. From these, \SI{10}{\micro\metre}-thick HOPG films were selected, cut to \SI{80}{\micro\metre} in diameter, and loaded into the sample chamber. Ruby powders were also added for pressure calibration~\cite{Dewaele_Ruby_2008}. Research-grade deionized water was then introduced into the chamber using a syringe. Microphotographs in the insets of Figs.~\ref{fig:fig3} and~\ref{fig:fig4} illustrate the sample space. Further details of the HOPG sample preparation in DACs are provided in Ref.~\cite{Lim_graphiteRaman_2016}.

Raman spectra were collected in a backscattering geometry at the GSECARS Raman Laboratory using a 2-W Innova 90P argon ion laser and a Raman spectrometer. This system allows for offline Raman spectroscopy and ruby fluorescence-based pressure measurements. A \SI{532}{\nano\metre} laser was used to excite the sample. Pressure was increased up to \SI{9.9}{\GPa} and subsequently released down to \SI{2.1}{\GPa}. A broad fluorescence background is also present, likely arising from nitrogen impurities in the type Ia diamonds, which fluoresce under \SI{532}{\nano\metre} laser excitation.

\section{Results}

\begin{figure}[b]
    \centering
    \includegraphics[width=\columnwidth]{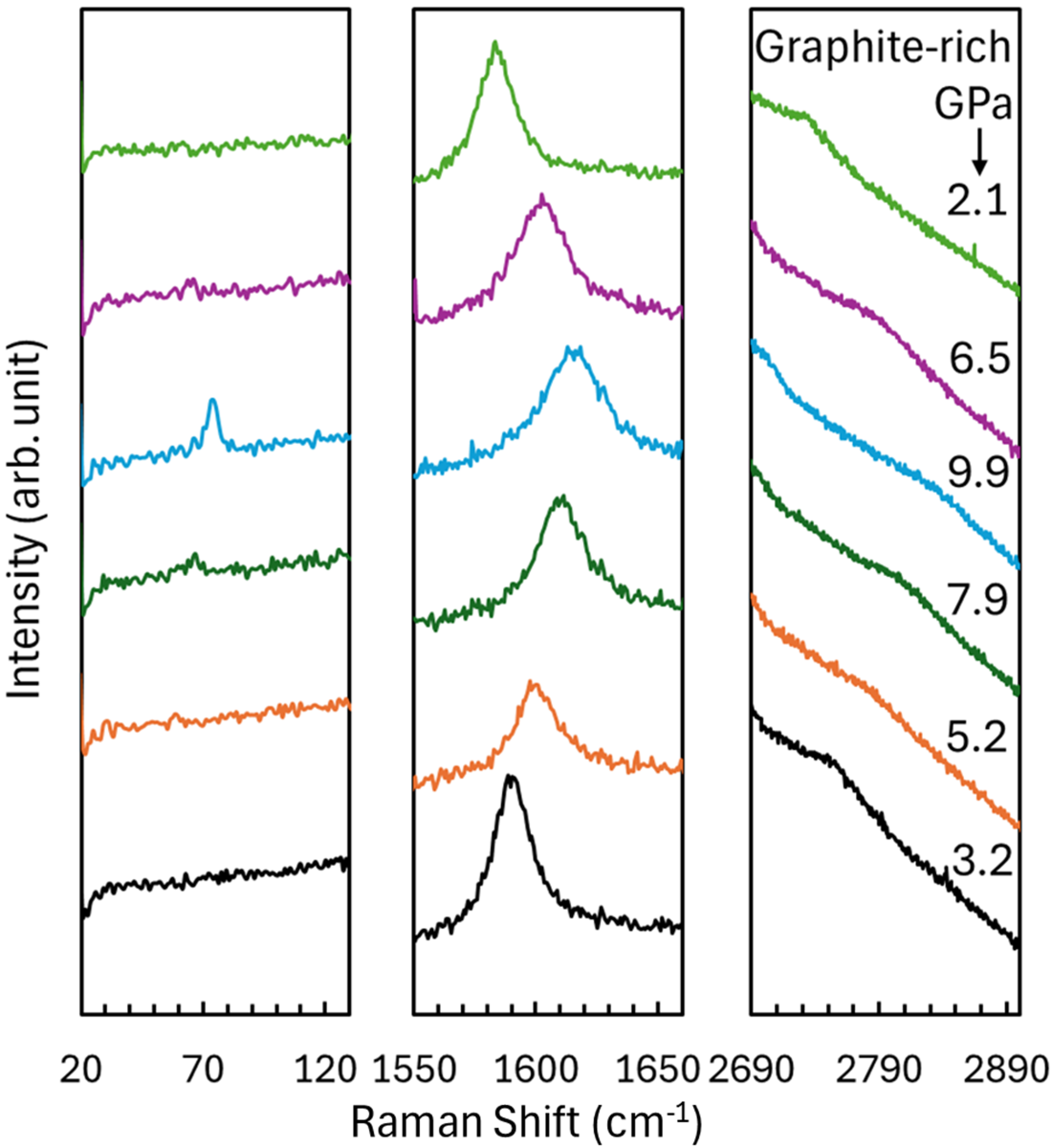}
    \caption{Selected regions of the Raman spectra of the graphite-rich area in a graphite–water mixture up to \SI{9.9}{GPa} (compression) and down to \SI{2.1}{GPa} (decompression), highlighting three characteristic features: the $E_{2\mathrm{g}}^{(1)}$ mode, the G band ($E_{2\mathrm{g}}^{(2)}$), and the 2D band, observed in the ranges of 20–\SI{130}{\per\centi\metre}, 1550–\SI{1660}{\per\centi\metre}, and 2690–\SI{2900}{\per\centi\metre}, respectively. A second-order Raman signal from the diamond anvils, appearing around \SI{2700}{\per\centi\metre}, precedes the 2D band. Sharp features with widths less than \SI{5}{\per\centi\metre} are considered spectral artifacts.}
    \label{fig:fig1}
\end{figure}

Figure~\ref{fig:fig1} presents the Raman spectra of the graphite-rich region in the graphite–water mixture, with water serving as the pressure medium, under compression up to \SI{9.9}{GPa} and decompression down to \SI{2.1}{GPa}. Three characteristic features are highlighted: the $E_{2\mathrm{g}}^{(1)}$ mode, the G band ($E_{2\mathrm{g}}^{(2)}$), and the 2D band, observed in the ranges of 20–\SI{130}{\per\centi\metre}, 1550–\SI{1660}{\per\centi\metre}, and 2690–\SI{2900}{\per\centi\metre}, respectively. Sharp peaks with widths narrower than \SI{5}{\per\centi\metre} are identified as spectral artifacts. The $E_{2\mathrm{g}}^{(1)}$ mode at $\sim$\SI{70}{\per\centi\metre} is relatively weak because it corresponds to shear motion between adjacent graphite planes and is challenging to detect due to its low frequency. Nevertheless, the high sensitivity of the Raman system allows this mode to be observed.

The G band, centered near \SI{1600}{\per\centi\metre}, corresponds to the in-plane stretching vibration of sp\textsuperscript{2} carbon atoms within the hexagonal lattice. It is a defining feature of graphene-related materials and is highly sensitive to doping, strain, and temperature~\cite{Wu_GrapheneRaman_2018}. The peak at $\sim$\SI{2800}{\per\centi\metre} corresponds to the 2D band, a second-order two-phonon overtone of the D band, although its signal is partially obscured by the diamond anvil Raman background (not shown). The 2D band is particularly sensitive to graphene layer number, and its shape and position can be used to assess the thickness of few-layer graphene~\cite{Wu_GrapheneRaman_2018}. With increasing pressure, all three Raman features shift to higher wavenumbers (blue shift), reflecting strengthened interatomic interactions under compression.

\begin{figure}[b]
    \centering
    \includegraphics[width=\columnwidth]{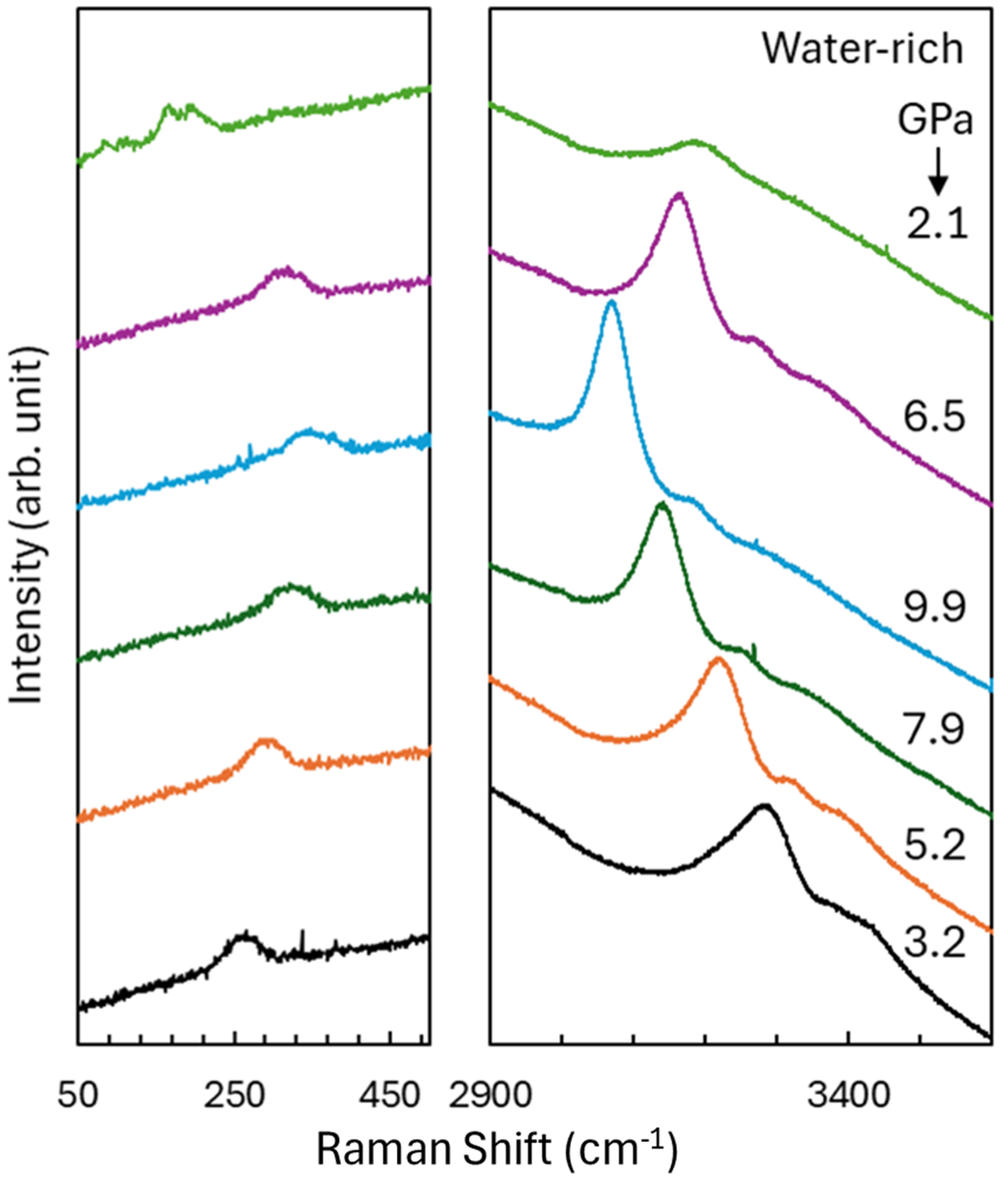}
    \caption{Selected regions of the Raman spectra of the water-rich area in a graphite–water mixture up to \SI{9.9}{GPa} (compression) and down to \SI{2.1}{GPa} (decompression), highlighting two characteristic regions: the librational band (50–\SI{500}{\per\centi\metre}) and the O–H stretching modes (2900–\SI{3660}{\per\centi\metre}), including the in-phase symmetric stretch ($A_{1\mathrm{g}}$), the in-phase asymmetric stretch ($E_{\mathrm{g}}$), and the out-of-phase stretch ($B_{1\mathrm{g}}$). Sharp features with widths less than \SI{5}{\per\centi\metre} are considered spectral artifacts.}
    \label{fig:fig2}
\end{figure}

Figure~\ref{fig:fig2} shows the Raman spectra of the water-rich region of the graphite–water mixture under the same pressure conditions. Two characteristic spectral regions are observed: the librational band (50–\SI{500}{\per\centi\metre}) and the O–H stretching region (2900–\SI{3660}{\per\centi\metre}). A librational band, arising from hindered rotational motion of water molecules constrained by hydrogen bonding or lattice interactions, appears near \SI{250}{\per\centi\metre}, consistent with earlier observations in ice VII~\cite{Pruzan_IceVIIRaman_1990}. The O–H stretching region consists of three modes: the in-phase symmetric stretch ($A_{1\mathrm{g}}$) at $\sim$\SI{3300}{\per\centi\metre}, the in-phase asymmetric stretch ($E_{\mathrm{g}}$) at $\sim$\SI{3400}{\per\centi\metre}, and the out-of-phase stretch ($B_{1\mathrm{g}}$) at $\sim$\SI{3450}{\per\centi\metre}. With increasing pressure, the librational band shifts to higher wavenumbers (blue shift), while the stretching modes shift to lower wavenumbers (red shift), reflecting enhanced hydrogen bonding and concomitant weakening of the O–H covalent bond.

\begin{figure}[b]
    \centering
    \includegraphics[width=\columnwidth]{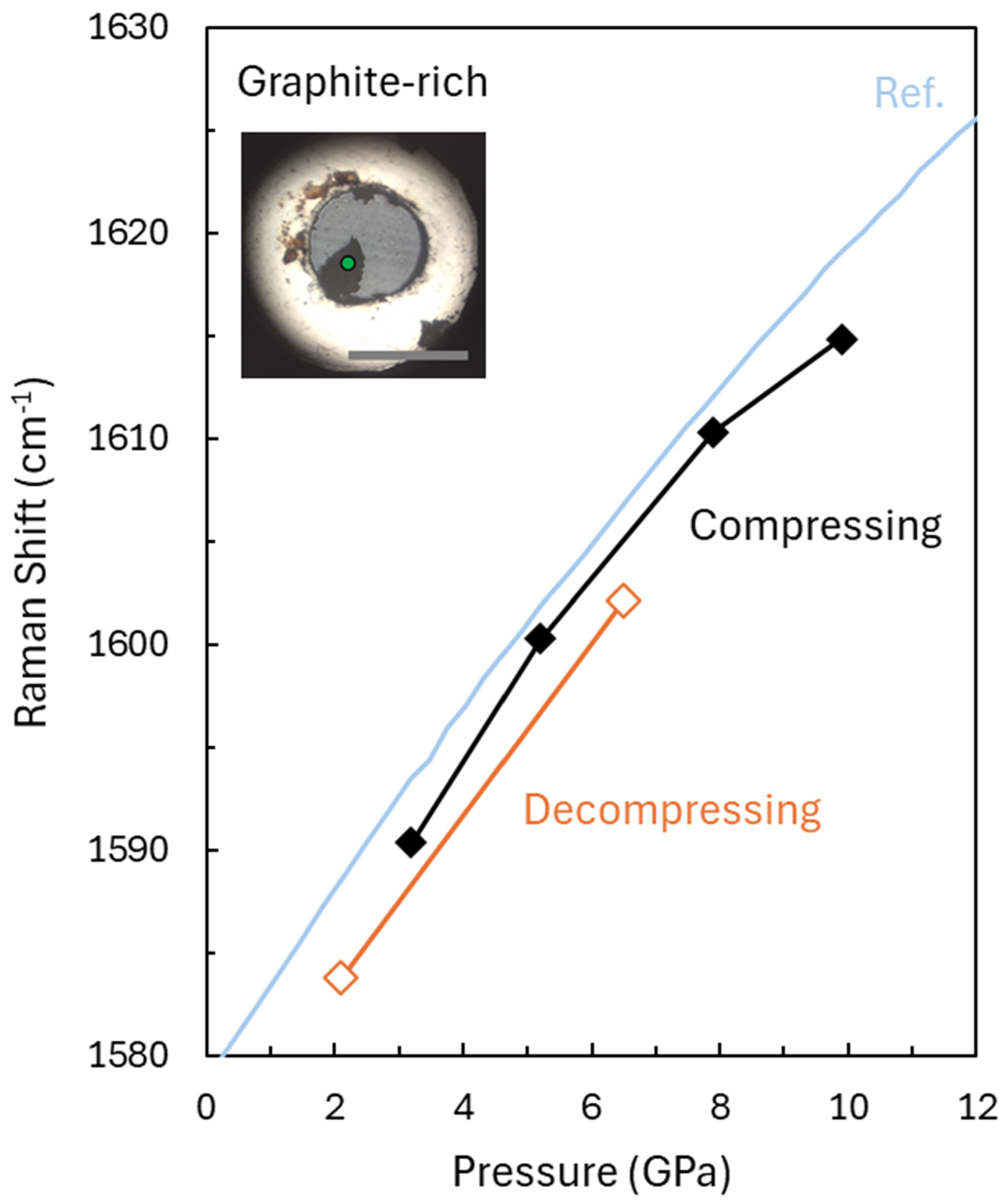}
    \caption{Raman shift of the G band ($E_{2\mathrm{g}}^{(2)}$) peak position in the graphite-rich region as a function of pressure, compared with literature values in light blue~\cite{Hanfland_GraphiteRaman_1989}. Black closed diamonds denote data obtained during compression, while orange open diamonds denote decompression. The inset shows a microphotograph of the mixture sample at \SI{3.2}{GPa}, with the green circle marking the laser-excited region on graphite (black area). The scale bar corresponds to \SI{150}{\micro\metre} in diameter.}
    \label{fig:fig3}
\end{figure}

Figure~\ref{fig:fig3} displays the pressure dependence of the Raman G band peak position of graphite in a water pressure medium, extracted from Fig.~\ref{fig:fig1}. The inset shows a microphotograph of the mixture at \SI{3.2}{GPa}, with the laser-excited graphite region (black area) marked by a green circle. The upward shift of the G band with pressure indicates that interlayer interactions between carbon atoms strengthen under compression. As pressure increases, graphite layers are forced closer together, enhancing restoring forces for in-plane vibrations, which produces the observed Raman shift. Comparison with earlier work (Ref.~\cite{Hanfland_GraphiteRaman_1989}), which used a methanol–ethanol pressure medium, shows overall good agreement. However, above \SI{8}{GPa}, the slope of the pressure dependence decreases relative to the earlier report, likely due to the influence of the water pressure medium. Water, as a pressure-transmitting medium, may alter the high-pressure response of graphite, potentially facilitating a lower-pressure transition from typical $sp^{2}$ bonding to a diamond-like $sp^{3}$ configuration~\cite{Tao_BilayerGraphene_2020}. Upon decompression, the G band shift is fully reversible.

\begin{figure}[b]
    \centering
    \includegraphics[width=\columnwidth]{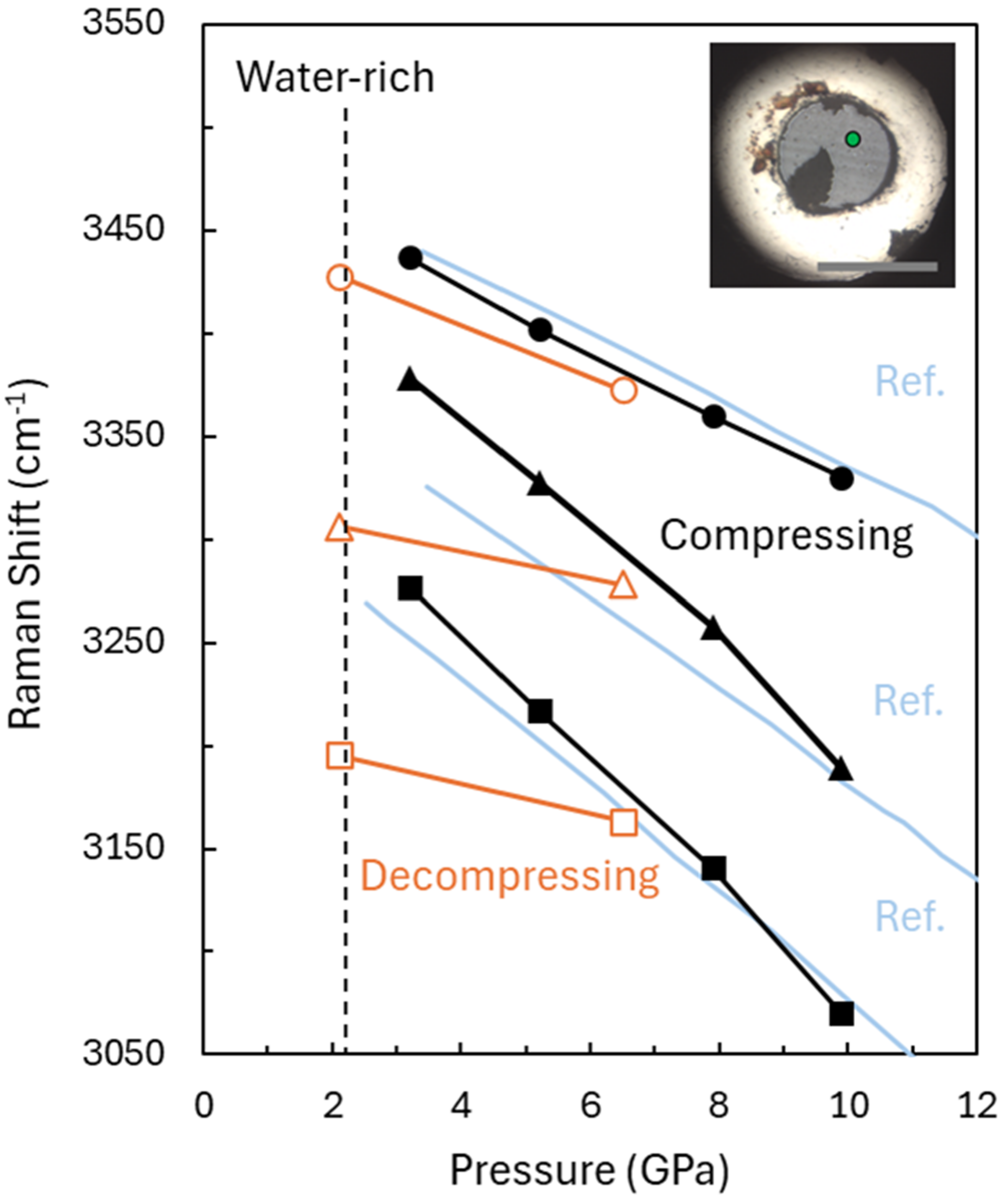}
    \caption{Raman shift of the O–H stretching modes (2900–\SI{3660}{\per\centi\metre}), including the in-phase symmetric stretch ($A_{1\mathrm{g}}$), the in-phase asymmetric stretch ($E_{\mathrm{g}}$), and the out-of-phase stretch ($B_{1\mathrm{g}}$), in the water-rich region as a function of pressure, compared with literature values shown in light blue~\cite{Pruzan_IceVIIRaman_1990}. Black closed symbols denote data obtained during compression, while orange open symbols denote decompression. The inset shows a microphotograph of the mixture sample at \SI{3.2}{GPa}, with the green circle marking the laser-excited region on water (gray area). The scale bar corresponds to \SI{150}{\micro\metre} in diameter.}
    \label{fig:fig4}
\end{figure}

Figure~\ref{fig:fig4} plots the Raman shifts of the three O–H stretching modes in the water-rich region—the $A_{1\mathrm{g}}$, $E_{\mathrm{g}}$, and $B_{1\mathrm{g}}$ modes—as a function of pressure, extracted from Fig.~\ref{fig:fig2}. The results are compared with previous work (Ref.~\cite{Pruzan_IceVIIRaman_1990}) and show good agreement. All three modes decrease in wavenumber with increasing pressure, consistent with strengthening intermolecular hydrogen bonding. Stronger hydrogen bonding weakens the O–H covalent bond, producing a red shift (lower energy). Upon decompression, a transition from ice VII to ice VI is indicated near \SI{2.2}{GPa}, marked by the dashed line~\cite{Hsieh_WaterRaman_2015}.

\section{Conclusions}
In summary, we have conducted a Raman spectroscopy of a graphite–water mixture under high pressure up to \SI{9.9}{GPa}. Three characteristic Raman features of graphite—the $E_{2\mathrm{g}}^{(1)}$ mode, the G band, and the 2D band—were clearly observed, along with the librational and O–H stretching modes of water. Both the graphite and water regions exhibited pressure-dependent Raman shifts that reflect the fundamental changes in intermolecular and interlayer interactions. The G band shifted upward with increasing pressure, consistent with strengthened interlayer coupling in graphite, while the O–H stretching modes shifted downward, indicating hydrogen-bond strengthening in compressed water. Notably, the G band response suggests that the presence of water as a pressure-transmitting medium may affect the high-pressure behavior of graphite, warranting further investigation at higher pressures.

\section*{Acknowledgments}
This work was supported by the Council on Faculty Research (CFR) at Eastern Illinois University, through both the Fall 2025 and Summer 2025 awards. The research was performed at GeoSoilEnviroCARS (University of Chicago, Sector 13) from the Advanced Photon Source, a U.S. Department of Energy (DOE) Office of Science user facility operated for the DOE Office of Science by Argonne National Laboratory under Contract No. DE-AC02-06CH11357. GeoSoilEnviroCARS is supported by the National Science Foundation–Earth Sciences via SEES: Synchrotron Earth and Environmental Science (EAR2223273). The work was supported by the Council on Faculty Research (CFR) at Eastern Illinois University, through both the Fall 2025 and Summer 2025 awards.

\bibliography{references}

\begin{thebibliography}{11}%
\makeatletter
\providecommand \@ifxundefined [1]{%
 \@ifx{#1\undefined}
}%
\providecommand \@ifnum [1]{%
 \ifnum #1\expandafter \@firstoftwo
 \else \expandafter \@secondoftwo
 \fi
}%
\providecommand \@ifx [1]{%
 \ifx #1\expandafter \@firstoftwo
 \else \expandafter \@secondoftwo
 \fi
}%
\providecommand \natexlab [1]{#1}%
\providecommand \enquote  [1]{``#1''}%
\providecommand \bibnamefont  [1]{#1}%
\providecommand \bibfnamefont [1]{#1}%
\providecommand \citenamefont [1]{#1}%
\providecommand \href@noop [0]{\@secondoftwo}%
\providecommand \href [0]{\begingroup \@sanitize@url \@href}%
\providecommand \@href[1]{\@@startlink{#1}\@@href}%
\providecommand \@@href[1]{\endgroup#1\@@endlink}%
\providecommand \@sanitize@url [0]{\catcode `\\12\catcode `\$12\catcode `\&12\catcode `\#12\catcode `\^12\catcode `\_12\catcode `\%12\relax}%
\providecommand \@@startlink[1]{}%
\providecommand \@@endlink[0]{}%
\providecommand \url  [0]{\begingroup\@sanitize@url \@url }%
\providecommand \@url [1]{\endgroup\@href {#1}{\urlprefix }}%
\providecommand \urlprefix  [0]{URL }%
\providecommand \Eprint [0]{\href }%
\providecommand \doibase [0]{https://doi.org/}%
\providecommand \selectlanguage [0]{\@gobble}%
\providecommand \bibinfo  [0]{\@secondoftwo}%
\providecommand \bibfield  [0]{\@secondoftwo}%
\providecommand \translation [1]{[#1]}%
\providecommand \BibitemOpen [0]{}%
\providecommand \bibitemStop [0]{}%
\providecommand \bibitemNoStop [0]{.\EOS\space}%
\providecommand \EOS [0]{\spacefactor3000\relax}%
\providecommand \BibitemShut  [1]{\csname bibitem#1\endcsname}%
\let\auto@bib@innerbib\@empty
\bibitem [{\citenamefont {Jackson}\ \emph {et~al.}(2005)\citenamefont {Jackson}, \citenamefont {Malba}, \citenamefont {Weir}, \citenamefont {Baker},\ and\ \citenamefont {Vohra}}]{Jackson_lanthanides_2005}%
  \BibitemOpen
  \bibfield  {author} {\bibinfo {author} {\bibfnamefont {D.~D.}\ \bibnamefont {Jackson}}, \bibinfo {author} {\bibfnamefont {V.}~\bibnamefont {Malba}}, \bibinfo {author} {\bibfnamefont {S.~T.}\ \bibnamefont {Weir}}, \bibinfo {author} {\bibfnamefont {P.~A.}\ \bibnamefont {Baker}},\ and\ \bibinfo {author} {\bibfnamefont {Y.~K.}\ \bibnamefont {Vohra}},\ }\href {https://doi.org/10.1103/PhysRevB.71.184416} {\bibfield  {journal} {\bibinfo  {journal} {Phys. Rev. B}\ }\textbf {\bibinfo {volume} {71}},\ \bibinfo {pages} {184416} (\bibinfo {year} {2005})}\BibitemShut {NoStop}%
\bibitem [{\citenamefont {Allan}\ \emph {et~al.}(2021)\citenamefont {Allan}, \citenamefont {Conejeros}, \citenamefont {Hart},\ and\ \citenamefont {Mohn}}]{Allan_EnergyLandscapes_2021}%
  \BibitemOpen
  \bibfield  {author} {\bibinfo {author} {\bibfnamefont {N.~L.}\ \bibnamefont {Allan}}, \bibinfo {author} {\bibfnamefont {S.}~\bibnamefont {Conejeros}}, \bibinfo {author} {\bibfnamefont {J.~N.}\ \bibnamefont {Hart}},\ and\ \bibinfo {author} {\bibfnamefont {C.~E.}\ \bibnamefont {Mohn}},\ }\href {https://doi.org/10.1007/s00214-021-02834-w} {\bibfield  {journal} {\bibinfo  {journal} {Theoretical Chemistry Accounts}\ }\textbf {\bibinfo {volume} {140}},\ \bibinfo {pages} {151} (\bibinfo {year} {2021})}\BibitemShut {NoStop}%
\bibitem [{\citenamefont {BOVENKERK}\ \emph {et~al.}(1959)\citenamefont {BOVENKERK}, \citenamefont {BUNDY}, \citenamefont {HALL}, \citenamefont {STRONG},\ and\ \citenamefont {WENTORF}}]{BOVENKERK_BundyDiamond_1959}%
  \BibitemOpen
  \bibfield  {author} {\bibinfo {author} {\bibfnamefont {H.~P.}\ \bibnamefont {BOVENKERK}}, \bibinfo {author} {\bibfnamefont {F.~P.}\ \bibnamefont {BUNDY}}, \bibinfo {author} {\bibfnamefont {H.~T.}\ \bibnamefont {HALL}}, \bibinfo {author} {\bibfnamefont {H.~M.}\ \bibnamefont {STRONG}},\ and\ \bibinfo {author} {\bibfnamefont {R.~H.}\ \bibnamefont {WENTORF}},\ }\href {https://doi.org/10.1038/1841094a0} {\bibfield  {journal} {\bibinfo  {journal} {Nature}\ }\textbf {\bibinfo {volume} {184}},\ \bibinfo {pages} {1094} (\bibinfo {year} {1959})}\BibitemShut {NoStop}%
\bibitem [{\citenamefont {Wang}\ \emph {et~al.}(2012)\citenamefont {Wang}, \citenamefont {Panzik}, \citenamefont {Kiefer},\ and\ \citenamefont {Lee}}]{Wang_graphiteRoomXRD_2012}%
  \BibitemOpen
  \bibfield  {author} {\bibinfo {author} {\bibfnamefont {Y.}~\bibnamefont {Wang}}, \bibinfo {author} {\bibfnamefont {J.~E.}\ \bibnamefont {Panzik}}, \bibinfo {author} {\bibfnamefont {B.}~\bibnamefont {Kiefer}},\ and\ \bibinfo {author} {\bibfnamefont {K.~K.~M.}\ \bibnamefont {Lee}},\ }\href {https://doi.org/10.1038/srep00520} {\bibfield  {journal} {\bibinfo  {journal} {Scientific Reports}\ }\textbf {\bibinfo {volume} {2}},\ \bibinfo {pages} {520} (\bibinfo {year} {2012})}\BibitemShut {NoStop}%
\bibitem [{\citenamefont {Tao}\ \emph {et~al.}(2020)\citenamefont {Tao}, \citenamefont {Du}, \citenamefont {Qi}, \citenamefont {Ni}, \citenamefont {Jiang},\ and\ \citenamefont {Zhu}}]{Tao_BilayerGraphene_2020}%
  \BibitemOpen
  \bibfield  {author} {\bibinfo {author} {\bibfnamefont {Z.}~\bibnamefont {Tao}}, \bibinfo {author} {\bibfnamefont {J.}~\bibnamefont {Du}}, \bibinfo {author} {\bibfnamefont {Z.}~\bibnamefont {Qi}}, \bibinfo {author} {\bibfnamefont {K.}~\bibnamefont {Ni}}, \bibinfo {author} {\bibfnamefont {S.}~\bibnamefont {Jiang}},\ and\ \bibinfo {author} {\bibfnamefont {Y.}~\bibnamefont {Zhu}},\ }\href {https://doi.org/10.1063/1.5135027} {\bibfield  {journal} {\bibinfo  {journal} {Applied Physics Letters}\ }\textbf {\bibinfo {volume} {116}},\ \bibinfo {pages} {133101} (\bibinfo {year} {2020})}\BibitemShut {NoStop}%
\bibitem [{\citenamefont {Dewaele}\ \emph {et~al.}(2008)\citenamefont {Dewaele}, \citenamefont {Torrent}, \citenamefont {Loubeyre},\ and\ \citenamefont {Mezouar}}]{Dewaele_Ruby_2008}%
  \BibitemOpen
  \bibfield  {author} {\bibinfo {author} {\bibfnamefont {A.}~\bibnamefont {Dewaele}}, \bibinfo {author} {\bibfnamefont {M.}~\bibnamefont {Torrent}}, \bibinfo {author} {\bibfnamefont {P.}~\bibnamefont {Loubeyre}},\ and\ \bibinfo {author} {\bibfnamefont {M.}~\bibnamefont {Mezouar}},\ }\href {https://doi.org/10.1103/PhysRevB.78.104102} {\bibfield  {journal} {\bibinfo  {journal} {Phys. Rev. B}\ }\textbf {\bibinfo {volume} {78}},\ \bibinfo {pages} {104102} (\bibinfo {year} {2008})}\BibitemShut {NoStop}%
\bibitem [{\citenamefont {Lim}\ and\ \citenamefont {Yoo}(2016)}]{Lim_graphiteRaman_2016}%
  \BibitemOpen
  \bibfield  {author} {\bibinfo {author} {\bibfnamefont {J.}~\bibnamefont {Lim}}\ and\ \bibinfo {author} {\bibfnamefont {C.-S.}\ \bibnamefont {Yoo}},\ }\href {https://doi.org/10.1063/1.4960733} {\bibfield  {journal} {\bibinfo  {journal} {Applied Physics Letters}\ }\textbf {\bibinfo {volume} {109}},\ \bibinfo {pages} {051905} (\bibinfo {year} {2016})},\ \Eprint {https://arxiv.org/abs/https://pubs.aip.org/aip/apl/article-pdf/doi/10.1063/1.4960733/14084351/051905\_1\_online.pdf} {https://pubs.aip.org/aip/apl/article-pdf/doi/10.1063/1.4960733/14084351/051905\_1\_online.pdf} \BibitemShut {NoStop}%
\bibitem [{\citenamefont {Wu}\ \emph {et~al.}(2018)\citenamefont {Wu}, \citenamefont {Lin}, \citenamefont {Cong}, \citenamefont {Liu},\ and\ \citenamefont {Tan}}]{Wu_GrapheneRaman_2018}%
  \BibitemOpen
  \bibfield  {author} {\bibinfo {author} {\bibfnamefont {J.-B.}\ \bibnamefont {Wu}}, \bibinfo {author} {\bibfnamefont {M.-L.}\ \bibnamefont {Lin}}, \bibinfo {author} {\bibfnamefont {X.}~\bibnamefont {Cong}}, \bibinfo {author} {\bibfnamefont {H.-N.}\ \bibnamefont {Liu}},\ and\ \bibinfo {author} {\bibfnamefont {P.-H.}\ \bibnamefont {Tan}},\ }\href {https://doi.org/10.1039/C6CS00915H} {\bibfield  {journal} {\bibinfo  {journal} {Chem. Soc. Rev.}\ }\textbf {\bibinfo {volume} {47}},\ \bibinfo {pages} {1822} (\bibinfo {year} {2018})}\BibitemShut {NoStop}%
\bibitem [{\citenamefont {Pruzan}\ \emph {et~al.}(1990)\citenamefont {Pruzan}, \citenamefont {Chervin},\ and\ \citenamefont {Gauthier}}]{Pruzan_IceVIIRaman_1990}%
  \BibitemOpen
  \bibfield  {author} {\bibinfo {author} {\bibfnamefont {P.}~\bibnamefont {Pruzan}}, \bibinfo {author} {\bibfnamefont {J.~C.}\ \bibnamefont {Chervin}},\ and\ \bibinfo {author} {\bibfnamefont {M.}~\bibnamefont {Gauthier}},\ }\href {https://doi.org/10.1209/0295-5075/13/1/014} {\bibfield  {journal} {\bibinfo  {journal} {Europhysics Letters}\ }\textbf {\bibinfo {volume} {13}},\ \bibinfo {pages} {81} (\bibinfo {year} {1990})}\BibitemShut {NoStop}%
\bibitem [{\citenamefont {Hanfland}\ \emph {et~al.}(1989)\citenamefont {Hanfland}, \citenamefont {Beister},\ and\ \citenamefont {Syassen}}]{Hanfland_GraphiteRaman_1989}%
  \BibitemOpen
  \bibfield  {author} {\bibinfo {author} {\bibfnamefont {M.}~\bibnamefont {Hanfland}}, \bibinfo {author} {\bibfnamefont {H.}~\bibnamefont {Beister}},\ and\ \bibinfo {author} {\bibfnamefont {K.}~\bibnamefont {Syassen}},\ }\href {https://doi.org/10.1103/PhysRevB.39.12598} {\bibfield  {journal} {\bibinfo  {journal} {Phys. Rev. B}\ }\textbf {\bibinfo {volume} {39}},\ \bibinfo {pages} {12598} (\bibinfo {year} {1989})}\BibitemShut {NoStop}%
\bibitem [{\citenamefont {Hsieh}\ and\ \citenamefont {Chien}(2015)}]{Hsieh_WaterRaman_2015}%
  \BibitemOpen
  \bibfield  {author} {\bibinfo {author} {\bibfnamefont {W.-P.}\ \bibnamefont {Hsieh}}\ and\ \bibinfo {author} {\bibfnamefont {Y.-H.}\ \bibnamefont {Chien}},\ }\href {https://doi.org/10.1038/srep08532} {\bibfield  {journal} {\bibinfo  {journal} {Scientific Reports}\ }\textbf {\bibinfo {volume} {5}},\ \bibinfo {pages} {8532} (\bibinfo {year} {2015})}\BibitemShut {NoStop}%
\end{thebibliography}%
\end{document}